\documentclass[12pt]{revtex4-1}
\usepackage{amsfonts,amssymb,amsmath,graphicx}
\begin{document}

\title{A method for calculating spectral statistics based on random-matrix universality with an application to the three-point correlations of the Riemann zeros}
\author{E. Bogomolny$^{1}$, J.P. Keating$^2$}
\affiliation{$^{1}$Univ. Paris-Sud, CNRS,
Laboratoire de Physique Th\'eorique et Mod\`eles statistiques, UMR8626,\\
Orsay, F-91405, France,\\
$^{2}$ School of Mathematics,\\
University of Bristol, Bristol, BS8 1TW, UK}

\begin{abstract}
We illustrate a general method for calculating spectral statistics that combines the universal (Random Matrix Theory limit) and the non-universal (trace-formula-related) contributions by giving a heuristic derivation of the three-point correlation function for the zeros of the Riemann zeta function.  The main idea is to construct a generalized Hermitian random matrix ensemble whose mean eigenvalue density coincides with a large but finite portion of the actual density of the spectrum or the Riemann zeros.  Averaging the random matrix result over remaining oscillatory terms related, in the case of the zeta function, to small primes leads to a formula for the three-point correlation function that is in agreement with results from other heuristic methods.  This provides support for these different methods.  The advantage of the approach we set out here is that it incorporates the determinental structure of the Random Matrix limit.        
\end{abstract}

\maketitle 

\section{Introduction}

Random matrix theory (RMT) was introduced in the 1950s to describe the highly-excited energy levels of heavy nuclei \cite{wigner_1}-\cite{bohigas}.  It has subsequently found numerous and important applications in many different branches of physics.   A review of various research directions in RMT is given, for example, in \cite{nina, RMTbook}.   Of all these directions, one is particularly unexpected: the use of RMT  in number theory.  Though statistical methods have many applications in number theory  (see e.g. \cite{kac}) the importance of RMT has attracted particular attention in recent years.  This line of investigation started with a theorem of Montgomery \cite{montgomery} relating to the pair correlation of the non-trivial zeros of the Riemann zeta function, and Dyson's remark that Montgomery's formula can be interpreted as saying that statistical properties of the zeros are the same as those of eigenvalues of large hermitian matrices with independent random elements.  Odlyzko's extensive numerical computations \cite{Odlyzko} provided compelling support for this conjecture, but also drew attention to the slow approach to the limit where RMT is expected to hold.  In the case of pair correlation, the approach to the limit was shown to be arithmetical in origin by Berry \cite{Berry} and a precise formula was derived in \cite{bogomolny_keating} which matches the numerical data extremely well \cite{SIAM}.  An analogous formula for the zeros of Dirichlet $L$-functions was recently derived in \cite{bogomolny_keating2}.   

Further developments in this direction came from a conjecture that the moments of the Riemann zeta function (as well other $L$-functions) can be calculated using the moments of characteristic polynomials in RMT \cite{keating, keating2}. Today there exists a large collection of conjectures which predict (in a good agreement with existing numerics) mean values of many different quantities related to number-theoretic zeta and $L$-functions using ideas from RMT.  For a review of the background to this area, see \cite{KS, KS2}.

Unfortunately, only a limited number of rigorous results have been obtained in this field.  Most conjectures are based on heuristic arguments which are very difficult (if not impossible) to justify mathematically. As different heuristics stress different points, it is of interest and importance to compare different methods of calculation.  It is also often the case that these heuristic methods lead to new insights into similar problems concerning the spectral statistics of quantum chaotic systems (see, e.g., \cite{bogomolny_keating, KM}).   

The purpose of this paper is to calculate a formula, like that derived in  \cite{bogomolny_keating}, for the three-point correlation function for the Riemann zeros using a method proposed in \cite{varenna}. This function has already been obtained in \cite{triple} by Conrey and Snaith using the ratio conjecture \cite{ratio}, which follows from heuristic manipulations of the approximate functional equation of the zeta function developed in \cite{CFKRS}. Our calculation was carried out independently of, and at the same time as \cite{triple} but we hesitated to publish the result mainly because both methods are general, permit to calculate, in principle, all correlation functions, and should lead to the same formulae. 

We decided to present our calculations now because there has recently been renewed interest in these kinds of formulae (see, e.g.~\cite{Zeev, CS}) and we believe that the approach we take here sheds significant new light on their structure.  We emphasize that this approach is completely different to the one involving the ratios conjecture.  It is based on the idea of exploiting random matrix universality, and it incorporates universal and non-universal (arithmetic) components in a novel way from the outset.  In particular, it combines the determinental structure of random matrix theory with the arithmetical terms in a way that appears more natural than in other approaches.  We see this as a significant advantage, because it is often a major difficulty to identify the combinatorial identities underpinning this structure \cite{nonlinearity, RudSar, Zeev, CS}.  It is also worth remarking that the formulae that emerge from this kind of approach  are useful for applications (cf.~\cite{nearest, DHKMS}).  And finally, there has recently been considerable focus on random matrix universality (see, e.g., \cite{RMTbook}), and our approach is likely to be of interest in that context too. 

The plan of the paper is the following. Section~\ref{rmt_universality} is devoted to a short discussion of random matrix universality. It is well known that standard random matrix ensembles with different one-body potentials lead to different (non-universal) densities of eigenvalues. Nevertheless, it is widely accepted that after unfolding,  local statistical properties of eigenvalues for all 'reasonable' ensembles are the same. This universality leads to an explicit expression for the random matrix kernel, conjecturally valid for any  mean density.  In Section~\ref{riemann} an expression for the density of Riemann zeros calculated by taking into account a large but finite number of prime numbers is presented. The primes entering this formula are chosen in such a way that they can be considered as independent under an average over a large window of heights on the critical axis. Inserting the finite expression for the Riemann zeros density into the random matrix kernel gives us formal correlation functions of Riemann zeros. But the result inevitably  has oscillations related with short primes. Averaging such oscillations over a large window leads to our main conjecture for correlation functions presented in Section~\ref{conjecture}. In Section~\ref{two_point} it is demonstrated how  it is possible to derive from this conjecture  the two-point correlation function of Riemann zeros. Section~\ref{three_point} contains an explicit calculation of the tree-point correlation function. The final expression agrees with the result of \cite{triple} obtained by a completely different method. Section~\ref{summary} is a brief summary of important formulae. 
 

\section{Random matrix universality}\label{rmt_universality}

It is well known (see e.g. \cite{mehta}) that the standard Gaussian Unitary Ensemble of $N\times N$ random matrices (GUE)  is determined as an ensemble of Hermitian matrices ($M_{mn}=M_{mn}^{\dag}$) whose elements  are random variables with the joint probability
\begin{equation}
P(M_{mn})=C_N \mathrm{e}^{-\mathrm{Tr}\, M^2}\prod_{j=1}^N \mathrm{d}M_{nn}
\prod_{1\leq m<n\leq N}\mathrm{d} \mathrm{Re}\,M_{mn}\ \mathrm{d} \mathrm{Im}\,M_{mn}
\label{gaussian}
\end{equation}    
where $C_N$ is a normalization constant. 

A natural generalization of this ensemble consists in choosing instead of  $\mathrm{Tr}\, M^2$ in the exponent an 'arbitrary' function $\mathrm{Tr}\, f(M)$ (called often the one-body potential) 
\begin{equation}
P(M_{mn})=C_N \mathrm{e}^{-\mathrm{Tr}\,f( M)}\prod_{j=1}^N \mathrm{d}M_{nn} 
\prod_{1\leq m<n\leq N}\mathrm{d} \mathrm{Re}\,M_{mn}\ \mathrm{d} \mathrm{Im}\,M_{mn}\ .
\label{ensemble_f}
\end{equation} 
All such ensembles permit one to integrate over angle-type variables and get the joint probability density for the eigenvalues $\lambda_j$ of the matrices $M_{mn}$ \cite{mehta}
\begin{equation}
P(\lambda_1,\lambda, \ldots, \lambda_N)=C^{\prime}_N \prod_{1\leq i<j\leq N}(\lambda_j-\lambda_i)^2 \exp \left (-\sum_{k=1}^Nf(\lambda_k) \right )\ .
\end{equation}
To calculate the $n$-point correlation function one has to fix $n$ eigenvalues $x_j=\lambda_j$ with $j=1,\ldots, n$ and to integrate over the remaining $N-n$ variables. For the ensemble considered, this can  be done by the method of orthogonal polynomials \cite{mehta}. One first introduces polynomials, $p_k(x)$ orthogonal with respect to the measure $\exp \left (-f(x) \right )$
\begin{equation}
\int \mathrm{e}^{-f(x) }p_k(x)p_r(x)\mathrm{d}x=\delta_{kr}\ .
\end{equation}
Then the $n$-point correlation function for the ensemble \eqref{ensemble_f} takes the form of the $n\times n$ determinant \cite{mehta}  
\begin{equation}
R(x_1,\ldots,x_n)=\det( K_N(x,y) )_{x,y=x_1,\ldots, x_n}
\label{n_corr_functions}
\end{equation}
where the kernel $K_N(x,y)$ is expressed through the orthogonal polynomials as follow
\begin{equation}
K_N(x,y)=\sum_{j=0}^{N-1} p_j(x)p_j(y)\mathrm{e}^{-f(x)/2-f(y)/2}\  .
\end{equation}
In particular, the mean level density, $\bar{\rho}(x)$, (i.e. one-point correlation function) is
\begin{equation}
\bar{\rho}(x)=K_N(x,x)\ .
\label{density}
\end{equation}
Usually one is interested in the limit $N\to\infty$ and the main question is what is then the limiting behaviour of this kernel. For the Gaussian ensemble \eqref{gaussian} the answer is well known \cite{mehta} since the orthogonal polynomials in this case  are  the usual Hermite polynomials
\begin{equation}
p_j(x)=\frac{1}{\sqrt{2^jj!\sqrt{\pi}}}H_j(x)\ . 
\end{equation} 
The mean level density \eqref{density} in this case is given by the famous "semicircle law", 
\begin{equation}
\bar{\rho}(x)=\left \{ \begin{array}{cc}\dfrac{1}{\pi}\sqrt{2N-x^2},& |x|<\sqrt{2N}\\0,&|x|>\sqrt{2N}\end{array}\right . \ ,
\end{equation} 
and there exists an explicit formula for the kernel $K_N(x,y)$ when $N$ is large. This formula takes an especially simple form in the bulk of the spectrum when $|x|,|y|\ll \sqrt{N}$
\begin{equation}
K_N(x,y)=\frac{\sin \pi \bar{\rho}(0)(x-y)}{\pi (x-y)}
\label{kernel_gue}
\end{equation}
where $\bar{\rho}(0)=\sqrt{2N}/\pi$ is the level density \eqref{density} at small $x$. 

For a general  one-body potential $f(x)$ the situation is more difficult.  The mean level density can be calculated for large $N$ from the Dyson equation \cite{mehta}
\begin{equation}
\mathrm{P}\int_{-\infty}^{\infty}\frac{\bar{\rho}(z)}{x-z}\mathrm{d}z=\tfrac{1}{2}f^{\prime}(x)
\label{dyson}
\end{equation}
where P indicates the principal value integral, but correlation functions are more difficult to obtain though    
there exists a vast literature on this subject (see e.g. \cite{orthogonal} and references therein). 

Instead of using rigorous asymptotic formulae, we shall argue as follows. It is well known that the mean level density is not a universal quantity. Different one-body potentials, $f(x)$, lead to different densities (cf. \eqref{dyson}). On the other hand, it is widely accepted that after unfolding all correlation functions in the local scale should be universal. The unfolding here means that one calculates statistical properties not of the true levels, $\lambda_n$, (which, in general, have a non-universal mean density $\bar{\rho}(x)$) but of new quantities 
\begin{equation}
e_n=\bar{N}(\lambda_n)
\end{equation}
where $\bar{N}(x)$ is the mean number of levels with $\lambda_j<x$, $\bar{N}(x)=\int^x \bar{\rho}(y) \mathrm{d}y$. 

By construction, the new levels, $e_n$, have unit mean density, and random matrix universality asserts that for these quantities the kernel has exactly the same form as in \eqref{kernel_gue}  with $\bar{\rho}(0)=1$
\begin{equation}
K(x,y)=\frac{\sin \pi (\bar{N}(x)-\bar{N}(y) )}{\pi (x-y)}\ .
\label{k_general}
\end{equation}
This formula is assumed  to be valid when (i) points $x$ and $y$ are far from the ends of the spectrum, and (ii) the mean number of levels $\bar{N}(x)$ is a smooth function of $x$ i.e. it changes slowly in the scale of the nearest levels. 

This expression is the concise manifestation of random matrix universality but we are  not aware that it has been proved in full generality. Nevertheless, it agrees with all we know (or conjecture) about universal behaviour of random matrix ensembles (at least for GUE types) and we shall apply  it below for the Riemann zeta function. 
\section{Riemann zeta function}\label{riemann} 
 
The Riemann zeta function is defined when Re$s>1$ as the sum over all integers (see e.g. \cite{riemann})
\begin{equation}
\zeta(s)=\sum_{n=1}^{\infty}\frac{1}{n^s}
\end{equation}    
or as the Euler product over prime numbers 
\begin{equation}
\zeta(s)=\prod_{p}\left (1-\frac{1}{p^s}\right )^{-1}\ .
\label{euler}
\end{equation}
It has an analytic continuation to the rest of the complex $s$-plane, except for a pole at $s=1$.  The celebrated Riemann Hypothesis states that all non-trivial zeros of this function, $\zeta(s_j)=0$, have the form
\begin{equation}
s_j=\tfrac{1}{2}+\mathrm{i}E_j
\end{equation}
with real $E_j$ (which may be thought of as analogous to quantum energies). 

Assuming the Riemann Hypothesis, it is easy to write down a formal expression for the density of these zeros (called in the mathematical literature Weil's explicit formula \cite{weil}). Indeed if one writes $\zeta(1/2+\mathrm{i}E)\sim \prod_j(E-E_j)$ then  the density of $E_j$ is
\begin{equation}
d(E)=-\frac{1}{\pi }\mathrm{Im}\,\frac{\partial }{\partial E}\ln \zeta \left (\tfrac{1}{2}+\mathrm{i}(E+\mathrm{i}\varepsilon)\right )_{\varepsilon\to +0}\ .
\end{equation}
Using the functional relation for the zeta function and \eqref{euler} one gets that, as usual, the density of zeros is a sum of two terms 
\begin{equation}
d(E)=\overline{d(E)} +d^{(\mathrm{osc})}(E) \ ,
\end{equation}
where $ \overline{d(E)}$ is the smooth part of the density, which, as $E\rightarrow\infty$ is given by 
\begin{equation}
 \overline{d(E)}\approx\frac{1}{2\pi}\ln \frac{E}{2\pi} \ ,
\end{equation}
and $d^{(\mathrm{osc})}(E)$ is the oscillating part of the density 
\begin{equation}
d^{(\mathrm{osc})}(E)=-\frac{1}{2\pi}\sum_{p}\sum_{n=1}^{\infty} 
\frac{\ln p}{ p^{n/2}}\left (\mathrm{e}^{\mathrm{i} n E\ln p}+\mathrm{e}^{-\mathrm{i} n E\ln p}\right )\ .
\label{osc}
\end{equation}
Of course, the sum over all primes $p$ diverges at real $E$ and this expression has no (clear) mathematical meaning (similar to all "physical" trace formulae).  It gains such a meaning when integrated against a sufficiently smooth test function.

Nevertheless, it is legitimate to write the 'true' density as a finite sum over primes with $p<p^*$ and an unknown remainder  related with large primes satisfying $p>p^*$
\begin{equation}
d(E)=\bar{\rho}(E)+\mathrm{large\;primes},\qquad \bar{\rho}(E)=\overline{d(E)} +\widetilde{d(E,p^*)}
\label{full_density}
\end{equation}
where $\widetilde{d(E,p^*)}$ is the same sum as in \eqref{osc} but taken over a finite set of primes with   $p<p^*$ (the value of $p^*$ will be chosen below)
\begin{equation}
\widetilde{d(E,p*)}=-\frac{1}{2\pi}\sum_{p<p*}\sum_{n=1}^{\infty} 
\frac{\ln p}{ p^{n/2}}\left (\mathrm{e}^{\mathrm{i} n E\ln p}+\mathrm{e}^{-\mathrm{i} n E\ln p}\right )=
\frac{1}{2\pi \mathrm{i}}\frac{\partial}{\partial E}\sum_{p<p^*}
\ln\frac{1-A_p\mathrm{e}^{\mathrm{i}\Phi_p(E)}}{1-A_p\mathrm{e}^{-\mathrm{i}\Phi_p(E)}}\ .  
\label{d}
\end{equation}
Here for further convenience  we introduce the notation
\begin{equation}
A_p=\frac{1}{\sqrt{p}},\qquad \Phi_p(E)=E\ln p \  .
\end{equation}
The knowledge of $\bar{\rho}(E)$ permits easily to calculate the mean number of levels corresponding to this density 
\begin{equation}
\bar{N}(x,p^*)\equiv \int^x_0 \bar{\rho}(E)\mathrm{d}E= \int^x_0  ( \overline{d(E)}+\widetilde{d(E,p^*)} )\mathrm{d}E\ .
\end{equation}
It is plain that 
\begin{equation}
\mathrm{e}^{2\pi \mathrm{i}  \bar{N}(E,p^*)}=\mathrm{e}^{2\pi \mathrm{i} \overline{N(E)}}\prod_{p<p^*}
\frac{1-A_p \mathrm{e}^{\mathrm{i}\Phi_p(E)} }{1-A_p\mathrm{e}^{-\mathrm{i}\Phi_p(E)} }
\label{N}
\end{equation}
 where 
\begin{equation}
\overline{N(E)}=\frac{E}{2\pi}\ln\frac{E}{2\pi \mathrm{e}}+\mathrm{const}\ .  
\end{equation}
For the Riemann zeta function the constant is known ($7/8$) but is irrelevant for our purpose.


\section{Main conjecture}\label{conjecture}

The principal point in the approach to statistical properties of Riemann zeros advocated here consists in the assumption that the large primes indicated in \eqref{full_density} give rise to GUE correlations \eqref{n_corr_functions} with random matrix kernel \eqref{k_general} where  $\bar{N}(E,p*)$ is determined by small primes \eqref{N}. Precisely,
  \begin{equation}
K(E_i,E_j)=\frac{\sin(\pi (\bar{N}(E_i,p*)-\bar{N}(E_j,p*)))}{\pi (E_i-E_j)}\  .
\label{main_kernel}
\end{equation}
Of course, in such an approach the exact mechanism by which large primes conspire to give this kernel is completely ignored. But as we shall show below this assumption permits us to calculate all low order terms for correlation functions of Riemann zeros in agreement with ones calculated by different mathods.
   
When $n$-point correlation functions are calculated from \eqref{n_corr_functions} using the kernel \eqref{main_kernel}, the result necessarily has oscillations related with oscillations in the "mean" density of zeros \eqref{d} produced by  short primes. 

Usually one is looking for statistical properties of a set of zeros close to a large value of $E$. In this case it is natural to write $E_j=E+e_j$ and then to average  around $E$. It means that we propose to calculate correlation functions of the Riemann zeros from the following expression 
\begin{equation}
R_n(e_1,e_2,\ldots, e_n)=\left \langle \left  \langle \det \Big ( K(E+e_i,E+e_j )\Big )_{i,j=1,\ldots,n}\right \rangle \right \rangle_{\Delta E} \ .
\label{R_n_definition}
\end{equation}
Here the average indicated by $\langle  \langle \ldots \rangle \rangle_{\Delta E}$ is to be  carried out over a large window of heights $E$
\begin{equation}
\langle  \langle F(E) \rangle \rangle_{\Delta E} \equiv \frac{1}{\Delta E}\int_{E-\Delta E/2}^{E+\Delta E/2} F(E^{\prime})\mathrm{d}E^{\prime} \ .
\end{equation}
Let us choose  the cut-off prime, $p^*$, and   the window, $\Delta E$, to fulfil  the  inequalities
\begin{equation}
1\ll p^*\ll\Delta E\ll E\ .
\label{inequalities}
\end{equation}
This choice permits one, at least formally, to calculate all necessary mean values. In particular,  one has
\begin{equation}
\langle  \langle \mathrm{e}^{\mathrm{i} n E\ln p } \rangle \rangle_{\Delta E} =0,\qquad \mathrm{for}\; p<p^*,\qquad n\in Z^*\ ,
\end{equation}
and
\begin{equation}
 \langle  \langle \mathrm{e}^{ \mathrm{i}  E(n_1\ln p_1 -n_2\ln p_2)} \rangle \rangle_{\Delta E} =\delta_{n_1,n_2} \delta_{p_1,p_2}\qquad \mathrm{for}\; p_1,p_2<p^*,\qquad n_1,n_2\in Z^* \ .
\label{orthogonality} 
\end{equation}
Therefore 
\begin{equation}
\langle  \langle  \tilde{d}(E,p^*) \rangle \rangle_{\Delta E} =0,\qquad \langle  \langle  \overline{d(E)} \rangle \rangle_{\Delta E} = \overline{d(E)}\ . 
\label{rho_average}
\end{equation}
More generally,  phases $E\ln p$ associated with different primes $p<p^*$ can be considered as independent random phases and the procedure of averaging a quasi-periodic function of these phases  is reduced to the integration over them
\begin{equation}
\left \langle \left \langle F\Big (\mathrm{e}^{\mathrm{i}E\ln p_1},\ldots, \mathrm{e}^{\mathrm{i}E\ln p_n} \Big ) \right \rangle \right\rangle_{\Delta E} 
=\int_0^{2\pi}\frac{\mathrm{d}\phi_1}{2\pi}\cdots \int_0^{2\pi}\frac{\mathrm{d}\phi_n}{2\pi}
F\Big (\mathrm{e}^{\mathrm{i}\phi_1},\ldots, \mathrm{e}^{\mathrm{i}\phi_n} \Big ) \ .
\label{good_average}
\end{equation}
Any averaging procedure for which this relation is fulfilled is suitable for our purposes and it can serve  as the definition of 'good' averaging.

Eq.~\eqref{R_n_definition} together with \eqref{main_kernel} and \eqref{N}  are our main formulae for correlation functions of zeros of the Riemann zeta function. In the next Sections we show how such formula can be used to calculate explicitly the two and three-point correlation functions. When performing the calculations we shall see that,  after averaging, the remaining terms can be divided into two groups. The first contains various sums over primes such that they have well defined values in the formal limit $p^*\to\infty$. The second, which includes divergent contributions, can be transformed to the  formally divergent product (with imaginary $s$)
\begin{equation}
f(s,p^*)= \prod_{p<p^*}\dfrac{1-p^{-1}}{1-p^{-1-s}}\ .
\end{equation}  
Under the  assumption  that
\begin{equation}
1 \ll \ln(p^*)\ll 1/|s|\;. 
\label{ll}
\end{equation}
This can be done as follows 
\begin{eqnarray}
& &f(s,p^*)=\lim_{t\to 0}\prod_{p<p^*}\frac{1-1/p^{1+t}}{1-1/p^{1+s}}
=\lim_{t\to 0}\frac{\zeta(1+s)}{\zeta(1+t)}
\prod_{p>p^*}\frac{1-1/p^{1+s}}{1-1/p^{1+t}}\nonumber \\
&\approx &\lim_{t\to 0}\frac{\zeta(1+s)}{\zeta(1+t)}
\exp (\int_{\ln(p^*)}^{\infty}\frac{\mathrm{d}u}{u}(\mathrm{e}^{-tu}-\mathrm{e}^{-su}))
=\lim_{t\to 0}\frac{\zeta(1+s)}{\zeta(1+t)}\exp \ln(s/t)=s\zeta(1+s)\ .
\end{eqnarray}
In the last step we use that according to our assumption $s\ln (p^*)\ll 1$ (with, of course, $t\ln(p^*)\ll 1$), and
\begin{equation}
\int_{0}^{\infty}\frac{\mathrm{d} u}{u}(\mathrm{e}^{-tu}-\mathrm{e}^{-su})=\ln s -\ln t\ . 
\end{equation}
(A more careful derivation can be given by using Eq.~3.14.1 of \cite{riemann}.)

This leads to the conclusion that under \eqref{ll} 
\begin{equation}
 \prod_{p<p^*}\frac{1-p^{-1}}{1-p^{-1-s}}\underset{p^*\to\infty}{\longrightarrow} s\zeta(1+s)
\label{zeta}
\end{equation} 
and we shall use this expression throughout the paper. 

\section{Two-point correlation function of Riemann zeros}\label{two_point}

The simplest non-trivial example of  a correlation function of Riemann zeros is the two-point correlation function. It was calculated in \cite{bogomolny_keating} by using the explicit form of the Hardy-Littlewood conjecture concerning the distribution of prime pairs. See \cite{bogomolny_keating2} for an extension to Dirichlet $L$-functions.  A formula identical to that obtained in \cite{bogomolny_keating} was shown also to follow from the ratios conjecture \cite{ratiosCS}.  Here we show that exactly the same result is obtained form Eq.~\eqref{R_n_definition} based on completely different assumptions. 

From  \eqref{R_n_definition} one gets
\begin{eqnarray}
R_2(e_1,e_2)&=& \left \langle \left  \langle \det \left ( \begin{array}{cc} K(E+e_1,E+e_1 ) & K(E+e_1,E+e_2 \\ K(E+e_2,E+e_1 )& K(E+e_2,E+e_2) \end{array} \right )\right  \rangle \right \rangle_{\Delta E} \nonumber\\
& =&  \langle  \langle \bar{\rho}(E+e_1)\bar{\rho}(E+e_2)\rangle \rangle_{\Delta E}  -
 \langle  \langle K^2(E+e_1,E+e_2) \rangle \rangle_{\Delta E} \ .
\end{eqnarray}
Here we use  $K(E,E)=\bar{\rho}(E)$ and $K(E_1,E_2)=K(E_2,E_1)$. It is worth remarking that in this approach we start with a determinant - in other approaches the main difficultly lies in identifying the combinatorial identities that match with a determinental form.  Noting that
\begin{eqnarray}
K(E_1,E_2)&=&\frac{\sin^2(\pi (\bar{N}(E_1,p*)-\bar{N}(E_2,p*)))}{\pi^2 (E_1-E_2)^2}\\
&=&-\frac{1}{4\pi^2(E_1-E_2)^2}\left (\mathrm{e}^{2\pi \mathrm{i} (\bar{N}(E_1,p^*)-\bar{N}(E_2,p^*)}-2+ \mathrm{e}^{-2\pi \mathrm{i} (\bar{N}(E_1,p^*)-\bar{N}(E_2,p^*)} \right ) \nonumber
\end{eqnarray}
and assuming the validity of \eqref{rho_average},  one gets 
\begin{equation}
R_{2}(e_1,e_2)=\overline{d(E)}^{\,2}+R_{2}^{\mbox{c}}(e_{1},e_2)\ .
\end{equation}
$R_{2}^{\mbox{c}}(e_{1},e_2)$ is the connected part of the two-point correlation function equals to the sum of two terms, the smooth term, $R_{2}^{\mbox{diag}}(e_1, e_2)$,  and the oscillatory term, $R_{2}^{\mbox{osc}}(e_1,e_2)$, 
\begin{eqnarray}
R_2^{\mbox{c}}(e_1,e_2)&\equiv & \langle  \langle K^2(E+e_1,E+e_2) \rangle \rangle_{\Delta E} =
R_{2}^{\mbox{diag}}(e_1, e_2)+R_{2}^{\mbox{osc}}(e_1,e_2) \ ,
\label{rtwo}
\end{eqnarray}
where
\begin{equation}
R_2^{\mbox{diag}}(e_1,e_2)= \langle \langle d_1\  d_2 \rangle \rangle_{\Delta E} -\frac{1}{2\pi^2\epsilon^2}
\label{R_diag}
\end{equation}
and
\begin{equation}
R_{2}^{\mbox{osc}}(e_1,e_2)=\frac{1}{4\pi^2 \epsilon^2} \langle \langle   \mathrm{e}^{2\pi \mathrm{i} (N_1-N_2)}+\mathrm{e}^{-2\pi \mathrm{i} (N_1-N_2)}\rangle \rangle_{\Delta E}\ .
\label{R_osc}
\end{equation}
Here and below we use the following notations: $ d_j=\widetilde{d(E+e_j, p*)}$, $N_j=\bar{N}(E+e_j,p*)$, and $\epsilon=e_1-e_2$. 

\subsection{Smooth terms}

The calculation of $\langle \langle d_1\  d_2 \rangle \rangle_{\Delta E}$ is straightforward and corresponds to the so-called diagonal approximation. From \eqref{d} one has 
\begin{eqnarray}
&4\pi^2 & \langle \langle d_1\  d_2 \rangle \rangle_{\Delta E} =\sum_{p_{1,2}<p*}\sum_{n_{1,2}=1}^{\infty} 
\frac{\ln p_1 \ln p_2}{p_1^{n_1/2}p_2^{n_2/2}} \Big \langle \Big \langle \left (\mathrm{e}^{\mathrm{i} n_1 (E+e_1)\ln p_1}+\mathrm{e}^{-\mathrm{i} n_1 (E+e_1)\ln p_1}\right ) \nonumber \\
&\times & \left (\mathrm{e}^{\mathrm{i} n_2 (E+e_2)\ln p_2}+\mathrm{e}^{-\mathrm{i} n_2 (E+e_2)\ln p_2}\right )\Big \rangle \Big \rangle_{\Delta E}
\nonumber\\
&=&\sum_{p<p*}\sum_{n=1}^{\infty}\frac{\ln^2 p}{p^{n} }
\Big (\mathrm{e}^{\mathrm{i} n \epsilon \ln p}+\mathrm{e}^{-\mathrm{i} n \epsilon\ln p}\Big )
= - \frac{\partial^2}{\partial \epsilon^2}\sum_{p<p*}\sum_{n=1}^{\infty}\frac{1}{n^2 p^{n} }
\Big(\mathrm{e}^{\mathrm{i} n \epsilon \ln p}+\mathrm{e}^{-\mathrm{i} n \epsilon\ln p}\Big )\nonumber\\ 
&=& - \frac{\partial^2}{\partial \epsilon^2}\sum_{p<p*}\sum_{n=2}^{\infty}\frac{1}{n^2 p^{n} }
\left (\mathrm{e}^{\mathrm{i} n \epsilon\ln p}+\mathrm{e}^{-\mathrm{i} n \epsilon\ln p}\right )
- \frac{\partial^2}{\partial  \epsilon^2}\sum_{p<p*}\frac{1}{ p }
\left (\mathrm{e}^{\mathrm{i} \epsilon \ln p}+\mathrm{e}^{-\mathrm{i} \epsilon\ln p}\right )
\label{d_1_d_2}
\end{eqnarray}
When $p^*\to\infty$ only the last term diverges. To calculate this it is convenient to use Eq.~\eqref{zeta}. By taking the logarithm of the both parts of this relation and  of its complex conjugate one obtains 
\begin{equation}
\sum_{p<p^*}\sum_{n=1}^{\infty}\frac{1}{n p^n}\left (\mathrm{e}^{\mathrm{i} n s \ln p}+\mathrm{e}^{-\mathrm{i}n s\ln p}\right )=
2\ln s +\ln |\zeta(1+\mathrm{i}s)|^2 + C
\end{equation}
with a constant $C=-2\ln \prod_{p<p^*}(1-1/p)$.  Consequently,
\begin{equation}
\sum_{p<p*}\frac{1}{ p }
\left (\mathrm{e}^{\mathrm{i} s\ln p}+\mathrm{e}^{-\mathrm{i} s\ln p}\right )=2\ln s +\ln |\zeta(1+\mathrm{i}s)|^2 -\sum_{p<p^*} \sum_{n=2}^{\infty}\frac{1}{n p^n}\left (\mathrm{e}^{\mathrm{i} n s \ln p}+\mathrm{e}^{-\mathrm{i}n s\ln p}\right )+ C\ .
\end{equation} 
Substituting this relation (with $s=\epsilon$) into \eqref{d_1_d_2} and taking into account the fact that the sums with $n\geq 2$ converge and $C$ is independent on $\epsilon$, we conclude that 
\begin{equation}
\langle \langle d_1\  d_2 \rangle \rangle_{\Delta E}=\frac{1}{2\pi^2 \epsilon^2}-\frac{1}{4\pi^2} \frac{\partial^2}{\partial \epsilon^2}\ln |\zeta(1+\mathrm{i} \epsilon )|^2 
-\frac{1}{4\pi^2} \frac{\partial^2}{\partial \epsilon^2}\sum_{p<p*}\sum_{n=2}^{\infty}\frac{1-n}{n^2 p^{n} }
\left (\mathrm{e}^{\mathrm{i} n \epsilon\ln p}+\mathrm{e}^{-\mathrm{i} n\epsilon \ln p}\right )\ .
\end{equation} 
Combining this expression and Eq.~\eqref{R_diag}, and using $\sum_{n=2}^{\infty}(1-n)x^n=-x^2/(1-x)^2$ gives 
\begin{equation}
R_{2}^{\mbox{diag}}(\epsilon)=-\frac{1}{4\pi^2}\frac{\partial^2}{\partial \epsilon^2}
\ln |\zeta(1+\mathrm{i}\epsilon)|^2-
\frac{1}{4\pi^2}\sum_p \ln^2 p\left( \frac{1}{(p^{1+\mathrm{i}\epsilon }-1)^2}+
\frac{1}{(p^{1-\mathrm{i} \epsilon}-1)^2} \right )\ .
\label{r_2diag}
\end{equation}

\subsection{Oscillatory terms}

The next step consists in calculating the oscillatory part of the two-point correlation function given by \eqref{R_osc}. Substituting  Eq.~\eqref{N} into \eqref{R_osc}, one gets 
\begin{equation}
\left \langle \left \langle  \mathrm{e}^{2\pi \mathrm{i} (N_1-N_2)} \right \rangle \right \rangle_{\Delta E} =
 \mathrm{e}^{2\pi \mathrm{i} \overline{d(E)}(e_1-e_2)}  \langle  \langle   R_p(E;e_1,e_2)  \rangle  \rangle_{\Delta E} 
\label{mean_exp_N} 
\end{equation}
where
\begin{equation}
R_p(E;e_1,e_2)= \prod_{p<p^*} \frac{[1-A_p \mathrm{e}^{-\mathrm{i}(\Phi_p(E)+e_2\ln p )}][1-A_p \mathrm{e}^{\mathrm{i}(\Phi_p(E)+e_1\ln p)}] }
{[1-A_p\mathrm{e}^{-\mathrm{i}(\Phi_p(E)+e_1\ln p)}][1-A_p \mathrm{e}^{\mathrm{i}(\Phi_p(E)+e_2\ln p)}] }\ .
\end{equation}
The averaging of $R_p(E,e_1,e_2)$ over $E$ can be done by using Eq.~\ref{good_average}. Therefore, the average over $E$ corresponds  to the independent integration  over phases $\Phi_p(E)=E\ln p_j$ 
\begin{equation}
\langle  \langle   R_p(E;e_1,e_2)  \rangle  \rangle_{\Delta E} =\prod_{p<p^*}  \langle \mathrm{R}_p (\Phi_p; e_1,e_2) \rangle_{\Phi_p} 
\end{equation}
where the average $\langle \mathrm{R}_p (\Phi_p; e_1,e_2) \rangle_{\Phi_p}$ is simply the mean value over all $\Phi_p$
\begin{equation}
\langle \mathrm{R}_p (\Phi_p; e_1,e_2) \rangle_{\Phi_p}=\frac{1}{2\pi}\int_0^{2\pi}\mathrm{R}_p (\Phi_p; e_1,e_2)\mathrm{d}\Phi_p
\  ,
\label{phi_average} 
\end{equation} 
and
\begin{equation}
\mathrm{R}_p (\Phi_p; e_1,e_2)=
\frac{[1-A_p \mathrm{e}^{-\mathrm{i}(\Phi_p +e_2\ln p)}][1-A_p \mathrm{e}^{\mathrm{i}(\Phi_p +e_1\ln p)} ] }
{[1-A_p\mathrm{e}^{-\mathrm{i}(\Phi_p +e_1\ln p)}][1-A_p \mathrm{e}^{\mathrm{i}(\Phi_p +e_2\ln p)}] }\ . 
\label{R_p}
\end{equation}
The calculation of the integral can conveniently be performed by complex integration. Putting $z=\mathrm{e}^{\mathrm{i}\Phi_p}$, one gets
\begin{equation}
\langle \mathrm{R}_p (\Phi_p;e_1,e_2) \rangle_{\Phi_p} =
 \frac{1}{2\pi} \oint \frac{[1-A_p z^{-1}\mathrm{e}^{-\mathrm{i}e_2\ln p }][1-A_p z  \mathrm{e}^{\mathrm{i}e_1\ln p}] }
{[1-A_p z^{-1} \mathrm{e}^{-\mathrm{i}e_1\ln p}][1-A_p z \mathrm{e}^{\mathrm{i}e_2\ln p}] }\frac{\mathrm{d}z}{z}
\end{equation}
where the integral is taken over the unit circle in the complex plane. 

As $A_p=p^{-1}<1$, inside the integration contour there are two poles. The first is at $z=0$ and the second at $z=A_p \mathrm{e}^{-\mathrm{i}e_1\ln p}$. Straightforward calculation gives
\begin{eqnarray}
\langle \mathrm{R}_p (\Phi_p;e_1,e_2) \rangle_{\Phi_p}&=&\mathrm{e}^{\mathrm{i}\epsilon \ln p}
+\frac{(1-A_p^2)(1-\mathrm{e}^{\mathrm{i}\epsilon \ln p })}{1-A_p^2\mathrm{e}^{-\mathrm{i}\epsilon \ln p }}=
\frac{1-2A_p^2+A_p^2\mathrm{e}^{\mathrm{i} \epsilon \ln p}}{1-A_p^2 \mathrm{e}^{-\mathrm{i}\epsilon \ln p }} \label{ww}\\
&=& \frac{(1-A_p^2)^2}{|1-A_p^2\, \mathrm{e}^{\mathrm{i}\epsilon \ln p }|^2}
\left ( 1-\frac{A_p^4(1-\mathrm{e}^{\mathrm{i}\epsilon \ln p })^2}{(1-A_p^2)^2} \right ) \nonumber \ ,
\label{average_R_p}
\end{eqnarray}
where as above $\epsilon=e_1-e_2$.

To calculate the product over $p<p^*$ we use Eq.~\eqref{zeta}; the final answer is   
\begin{equation}
R_{2}^{\mbox{osc}}(\epsilon)=\frac{1}{4\pi^2}\mathrm{e}^{2\pi \mathrm{i} \overline{d(E)}\epsilon}|\zeta(1+\mathrm{i}\epsilon)|^2
\prod_p \left (1-\frac{(1-p^{\mathrm{i}\epsilon })^2}{(p-1)^2}\right ) + \mbox{ c.c. }\ .
\label{r_2osc}
\end{equation}
Expressions \eqref{r_2diag} and \eqref{r_2osc} constitute two parts of the connected two-point correlation function of zeros of the Riemann zeta function. It is important  to stress that the same expressions were obtained in \cite{bogomolny_keating} by a completely different method based on the Hardy-Littlewood conjecture concerning the distribution of prime pairs.  Exactly the same formulae are also derived from the ratio conjecture \cite{ratiosCS}. Of course, all these methods are heuristic and cannot be considered as a true mathematical proof. Nevertheless, their mutual agreement means that if such a formula exists, it is likely to be given by the above expressions.   

It is plain that when $\epsilon\to 0$, Eq.~\ref{r_2osc}  reproduces the two-point correlation function for the GUE ensemble of random matrices
\begin{equation}
R_2^{\mbox{c}}(\epsilon)\underset{\epsilon\to 0}{\longrightarrow} -\frac{\sin^2(\pi \bar{d}\epsilon)}{\pi^2 \epsilon^2}\  .
\end{equation}
In \cite{nonlinearity} using an averaged version of the Hardy-Littlewood conjecture it was shown that at small separations all correlation functions of the Riemann zeros agree with random matrix predictions.  

\section{Three-point correlation function}\label{three_point}

The purpose of this Section is to calculate explicitly the three-point correlation function of zeros of the Riemann zeta function using the method discussed in the previous Sections. 

By definition 
\begin{equation}
R_3(e_1,e_2,e_3)=\left \langle \left \langle  \left | 
\begin{array}{ccc}
\bar{\rho}_1&K_{12}&K_{13}\\
K_{21}&\bar{\rho}_2&K_{23}\\
K_{31}&K_{32}&\bar{\rho}_3
\end{array}\right | \right  \rangle \right  \rangle_{\Delta E} 
\end{equation}
where $\bar{\rho}_j=\rho(E+e_j)$ and $K_{ij}=K(E+e_i, E+e_j)$.

Expanding the determinant one gets
\begin{equation}
R_3(e_1,e_2,e_3)=\bar{\rho}_1\, \bar{\rho}_2\, \bar{\rho}_3+K_{12}\, K_{23}\, K_{31}+K_{21}\, K_{13}\, K_{32}-
\bar{\rho}_1\,  K_{23}K_{32}-\bar{\rho}_2\,  K_{13}\, K_{31}-\bar{\rho}_3 \, K_{12}\, K_{21} \ .
\end{equation}
Each $K_{ij}$ is given by \eqref{main_kernel} and it is straightforward to check that
\begin{equation}
K_{12}\, K_{23}\, K_{31}=\frac{1}{(2\pi \mathrm{i})^3 e_{12} e_{23 }e_{31}}
\left (
\mathrm{e}^{2\pi \mathrm{i} (N_1-N_2)}+\mathrm{e}^{2\pi \mathrm{i} (N_2-N_3)}+\mathrm{e}^{2\pi \mathrm{i} (N_3-N_1)}-
\mathrm{  c.c. }\right ) 
\end{equation}
with $e_{ij}=e_i-e_j$.

From \eqref{full_density}  $\bar{\rho}_j\equiv \bar{\rho}(E+e_j)$ is the sum of two terms, $\bar{\rho}_j=\overline{d(E+e_j)} +\widetilde{d(E+e_j,p^*)}$. Using $\overline{d(E+e_j)}=\overline{d(E)}+ \mathcal{O}(e_j/E)$, ignoring the last correction terms,  and, as above, denoting $d_j=\widetilde{d(E+e_j,p^*)}$  one obtains
\begin{equation}
R_3(e_1,e_2,e_3)=\overline{d(E)}^{\ 3}+\overline{d(E)} \, R_2^{\mbox{c}}(e_{12})+
\overline{d(E)} \, R_2^{\mbox{c}}(e_{23})+
\overline{d(E)} \, R_2^{\mbox{c}}(e_{31}) +R_{3}^{\mbox{c}}(e_1,e_2,e_3)\ .
\end{equation}
Here $R_2^{\mbox{c}}(e_{ij})$ is the connected two-point correlation function \eqref{rtwo}  calculated in the previous Section, and   $R_{3}^{\mbox{c}}(e_1,e_2,e_3)$ is the connected three-point function conveniently written as the sum of two terms  
\begin{equation}
R_{3}^{\mbox{c}}(e_1,e_2,e_3)= R_{3}^{\mbox{diag}}(e_1,e_2,e_3)+R_{3}^{\mbox{osc}}(e_1,e_2,e_3) .
\label{rthree}
\end{equation}
Here $R_{3}^{\mbox{diag}}(e_1,e_2,e_3)$ is a smooth part 
\begin{equation}
R_{3}^{\mbox{diag}}(e_1,e_2,e_3)=\langle \langle d_1 d_2 d_3 \rangle \rangle_{\Delta E}\ ,
\label{diag}
\end{equation}
and $R_{3}^{\mbox{osc}}(e_1,e_2,e_3)$ is an oscillatory part
\begin{equation}
R_{3}^{\mbox{osc}}(e_1,e_2,e_3)=-\frac{1}{(2\pi i)^3}\left ( 
\frac{r_{12 }}{e_{12}^2}+
\frac{r_{23}}{e_{23}^2}+
\frac{r_{31}}{e_{31}^2}+\mbox{ c.c. }\right )
\end{equation}
where 
\begin{eqnarray}
r_{12}&=&\left \langle \left \langle \left [ 2\pi \mathrm{i} 
d_3-\frac{2 e_{12}}{e_{23}e_{31}}\right ]\mathrm{e}^{2\pi \mathrm{i} (N_1-N_2)}\right \rangle \right \rangle_{\Delta E} \ ,\\
r_{23}&=&\left \langle \left \langle \left [ 2\pi \mathrm{i} 
d_1-\frac{2 e_{23}}{e_{12}e_{31}}\right ]\mathrm{e}^{2\pi \mathrm{i} (N_2-N_3)} \right \rangle \right \rangle_{\Delta E}\ ,\\
r_{31}&=&\left \langle \left \langle \left [ 2\pi \mathrm{i} 
d_2 -\frac{2e_{31}}{e_{12}e_{23}}\right ]\mathrm{e}^{2\pi \mathrm{i} (N_3-N_1)}\right \rangle \right \rangle_{\Delta E}\  . 
\end{eqnarray}

\subsection{Smooth terms}

The calculation of smooth (diagonal) contributions \eqref{diag} for the three-point correlation function of Riemann zeros is simplified by the fact that after averaging terms divergent when $p^*\to \infty$  disappear and only convergent sums remain.  
Indeed, the average over $E$  removes all products with different primes (cf. \eqref{orthogonality}). Therefore 
\begin{eqnarray}
&&R_{3}^{\mbox{diag}}(e_1,e_2,e_3)\equiv  \langle \langle d_1\, d_2\,  d_3\rangle \rangle_{\Delta E} =-\frac{1}{(2\pi)^3}\sum_p \ln^3 p \sum_{n_1,n_2,n_3=1}^{\infty}A_p^{n_1+n_2+n_3} \nonumber\\ 
&\times & 
\Big \langle 
\left (\mathrm{e}^{\mathrm{i} n_1(\Phi_p(E)+e_1\ln p)}+\mathrm{e}^{-\mathrm{i} n_1(\Phi_p(E)+e_1\ln p)}\right)
\left (\mathrm{e}^{\mathrm{i} n_2(\Phi_p(E)+e_2\ln p)}+\mathrm{e}^{-\mathrm{i} n_2(\Phi_p(E)+e_2\ln p)}\right)\nonumber\\
&\times &\left(\mathrm{e}^{\mathrm{i} n_3(\Phi_p(E)+e_3\ln p)}+\mathrm{e}^{-\mathrm{i} n_3(\Phi_p(E)+e_3\ln p)}\right )
 \Big \rangle_{\Phi_p}\  .
\end{eqnarray} 
After averaging over $\Phi_p$, non-zero result gives only diagonal terms with $n_1=n_2+n_3$, $n_2=n_1+n_3$, and $n_3=n_1+n_2$. So 
\begin{eqnarray}
&&R_{3}^{\mbox{diag}}(e_1,e_2,e_3)= -\frac{1}{(2\pi)^3}\sum_p \ln^3 p \left [
\frac{1}{(p^{1-\mathrm{i} e_{12}}-1)(p^{1-\mathrm{i} e_{13}}-1)}+\right . \nonumber\\
&&+\left .\frac{1}{(p^{1-\mathrm{i} e_{21}}-1)(p^{1-\mathrm{i} e_{23}}-1)}
+ \frac{1}{(p^{1-\mathrm{i} e_{32}}-1)(p^{1-\mathrm{i} e_{31}}-1)}\right ]+\mathrm{c.c.}\ . 
\label{rdiag}
\end{eqnarray}
\subsection{Oscillatory terms}

Using Eqs.~(\ref{N}) and (\ref{d}) we obtain 
\begin{eqnarray}
&&  \left \langle \left \langle   2\pi \mathrm{i} d_3 \mathrm{e}^{2\pi \mathrm{i} (N_1-N_2)}\right \rangle \right \rangle_{\Delta E} \nonumber\\
&&=\mathrm{e}^{2\pi \mathrm{i} \overline{d(E)} e_{12}}\frac{\partial}{\partial e_3}
\left \langle \left \langle  \left [ \sum_{p<p^*} 
\ln  \frac{1-A_p\mathrm{e}^{\mathrm{i}\Phi_p(3)}}{1-A_p\mathrm{e}^{-\mathrm{i}\Phi_p(3)}} \right ] 
\prod_{p<p^*} \frac{(1-A_p\mathrm{e}^{-\mathrm{i}\Phi_p(2)})(1-A_p \mathrm{e}^{\mathrm{i}\Phi_p(1)})}
{(1-A_p\mathrm{e}^{-\mathrm{i}\Phi_p(1)})(1-A_p\mathrm{e}^{\mathrm{i}\Phi_p(2)})} \right \rangle \right \rangle_{\Delta E}  
\label{sum_product}
\end{eqnarray}
where $\Phi_p(j)\equiv \Phi_p(E+e_j)=\Phi_p(E)+e_p\ln p$.

According to Eq.~\eqref{good_average} the average over $E$ is equivalent to the mean value over all phases. In Eq~\eqref{sum_product} the sum over primes is multiplied by the product over the same primes. Therefore the average of each term in the sum, say with $p=q$,  reduces to the following product of the averages
\begin{eqnarray}
& &\left \langle \left \langle \frac{\partial}{\partial e_3}
\ln \left [ \frac{1-A_q\mathrm{e}^{\mathrm{i}\Phi_q(3)}}{1-A_q\mathrm{e}^{-\mathrm{i}\Phi_q(3)}} \right ]  
\prod_{p<p^*} \frac{(1-A_p\mathrm{e}^{-\mathrm{i}\Phi_p(2)})(1-A_p \mathrm{e}^{\mathrm{i}\Phi_p(1)})}
{(1-A_p\mathrm{e}^{-\mathrm{i}\Phi_p(1)})(1-A_p\mathrm{e}^{\mathrm{i}\Phi_p(2)})} \right \rangle \right \rangle_{\Delta E}\nonumber \\
&=&  T_q(e_1,e_2,e_3)\ \prod_{p\neq q} \langle R_p(\Phi_p ; e_1,e_2)\rangle_{\Phi_p} 
\end{eqnarray}
where
\begin{eqnarray}
T_q( e_1,e_2,e_3)&=&\int_0^{2\pi} \frac{\mathrm{d}\Phi_q}{2\pi}  \frac{\partial}{\partial e_3}
\ln  \left [\frac{1-A_q\mathrm{e}^{\mathrm{i}(\Phi_q+e_3\ln q) }}{1-A_q\mathrm{e}^{-\mathrm{i}(\Phi_q+e_3\ln p)}}\right ]\nonumber\\
&\times &  
\frac{(1-A_q\mathrm{e}^{-\mathrm{i}(\Phi_q+e_2\ln q)})(1-A_q \mathrm{e}^{\mathrm{i}(\Phi_q+e_1\ln q)})}
{(1-A_q\mathrm{e}^{-\mathrm{i}(\Phi_q+e_1\ln q)})(1-A_q\mathrm{e}^{\mathrm{i}(\Phi_q+e_2\ln q)})}\ ,  
\end{eqnarray}
with  $R_p(\Phi_p ; e_1,e_2)$  given by Eq.~\eqref{R_p}, and its average by Eq.~\eqref{average_R_p}.

As in the previous Section it is convenient to  calculate $T_q(\Phi_q; e_1,e_2,e_3)$ by complex integration. Denoting $z=\mathrm{e}^{\mathrm{i}(\Phi_q}$, $A_q=A$, and $f_j=e_j\ln q$ one has
 \begin{equation}
  T_q( e_1,e_2,e_3) = - \frac{\ln q}{2\pi}\oint \left  [\frac{Az\mathrm{e}^{\mathrm{i}f_3} }{1-Az\mathrm{e}^{\mathrm{i}f_3}} + \frac{A\mathrm{e}^{-\mathrm{i}f_3}}{z-A\mathrm{e}^{-\mathrm{i}f_3}} \right ]  \frac{(1-A z \mathrm{e}^{\mathrm{i} f_1})(z-A \mathrm{e}^{-\mathrm{i}f_2})}{(z-A\mathrm{e}^{-\mathrm{i}f_1})(1-Az\mathrm{e}^{\mathrm{i}f_2})}\frac{\mathrm{d}z}{z}\ .
\end{equation}
Inside the unit circle the integrand has 3 poles, $z=0$, $z=A\mathrm{e}^{-\mathrm{i}f_1}$, and $z=A\mathrm{e}^{-\mathrm{i}f_3}$. 
Direct calculations give ($\phi_{ij}=f_i-f_j$)
\begin{eqnarray}
&&T_q( e_1,e_2,e_3) = -\mathrm{i}\ln q  \left  [- \mathrm{e}^{\mathrm{i}\phi_{12}} +\Big (  
\frac{A^2 \mathrm{e}^{\mathrm{i}\phi_{31}} }{1-A^2 \mathrm{e}^{\mathrm{i}\phi_{31}}}+
\frac{1}{\mathrm{e}^{\mathrm{i}\phi_{31}}-1} \Big ) 
\frac{(1-A^2)(1-\mathrm{e}^{\mathrm{i}\phi_{12}})}{1-A^2\mathrm{e}^{\mathrm{i}\phi_{21}}} \right .\nonumber \\ 
&&+ \left . \frac{(1-A^2\mathrm{e}^{\mathrm{i}\phi_{13}})(1-\mathrm{e}^{\mathrm{i}\phi_{32}})}{(1-\mathrm{e}^{\mathrm{i}\phi_{31}})(1-A^2\mathrm{e}^{\mathrm{i}\phi_{23}})}\right ]
= -\mathrm{i}\ln q   \frac{(1-A^2)(1-\mathrm{e}^{\mathrm{i}\phi_{12}})}{1-A^2\mathrm{e}^{\mathrm{i}\phi_{21}}}\left  [
\frac{A^2 \mathrm{e}^{\mathrm{i}\phi_{31}}}{1-A^2 \mathrm{e}^{\mathrm{i}\phi_{31}}}+\frac{A^2 \mathrm{e}^{\mathrm{i}\phi_{23}}}{1-A^2 \mathrm{e}^{\mathrm{i}\phi_{23}}}\right ] \ .
\end{eqnarray}
Re-introducing the full notation it follows that ($e_{ij}=e_i-e_j$)
\begin{equation}
T_q( e_1,e_2,e_3)=\frac{\partial}{\partial e_3} \ln \left  [ \frac{1-A_q^2 \mathrm{e}^{\mathrm{i}e_{31}\ln q}}{1-A_q^2 \mathrm{e}^{\mathrm{i}e_{23}\ln q}} \right ] \frac{(1-A_q^2)(1-\mathrm{e}^{\mathrm{i}e_{12}\ln q})}{1-A_q^2\mathrm{e}^{\mathrm{i}e_{21}\ln q}} \ .
\end{equation}
Using Eq.~\eqref{ww} one gets
\begin{eqnarray}
r_{12}&=&\mathrm{e}^{2\pi \mathrm{i} \overline{d(E)}e_{12}}\prod_{p<p^*}
\frac{(1-A_p^2)^2}{|1-A_p^2 \mathrm{e}^{\mathrm{i}e_{12}\ln p}|^2} 
\left (1-\frac{A_p^4}{(1-A_p^2)^2}(1-\mathrm{e}^{\mathrm{i}e_{12}\ln p} )^2\right  )\times \\
&\times &\frac{\partial}{\partial e_3}\left [\sum_{q<p^*} 
\frac{(1-A_q^2)(1-\mathrm{e}^{\mathrm{i}e_{12}\ln q})}{1-2A_q^2+A_q^2 \mathrm{e}^{\mathrm{i}e_{12}\ln q}} 
\ln \frac{1-A_q^2\mathrm{e}^{\mathrm{i}e_{31}\ln q}}{1-A_q^2\mathrm{e}^{\mathrm{i}e_{23}\ln q}}+ \ln \frac{e_{31}^2}{e_{32}^2} \right ] \ .
 \end{eqnarray}
The summand in the square brackets can be transformed as follows
\begin{eqnarray}
&&\frac{(1-A^2)(1-\mathrm{e}^{\mathrm{i}\phi_{12}})}{1-2A^2+A^2\mathrm{e}^{\mathrm{i}\phi_{12}}} 
 \ln \frac{1-A^2\mathrm{e}^{\mathrm{i}\phi_{31}}}{1-A^2\mathrm{e}^{\mathrm{i}\phi_{23}}}=
\ln \frac{1-A^2\mathrm{e}^{\mathrm{i}\phi_{31}}}{1-A^2\mathrm{e}^{\mathrm{i}\phi_{23}}}+\ln \frac{1-A^2\mathrm{e}^{-\mathrm{i}\phi_{31}}}{1-A^2\mathrm{e}^{-\mathrm{i}\phi_{23}}} +\nonumber \\
&&+\frac{A^2-\mathrm{e}^{\mathrm{i}\phi_{12}}}{1-2A^2+A^2\mathrm{e}^{\mathrm{i}\phi_{12}}}\ln \frac{1-A^2\mathrm{e}^{\mathrm{i}\phi_{31}}}{1-A^2\mathrm{e}^{\mathrm{i}\phi_{23}}}-\ln \frac{1-A^2\mathrm{e}^{-\mathrm{i}\phi_{31}}}{1-A^2\mathrm{e}^{-\mathrm{i}\phi_{23}}}=\nonumber \\
&&=\ln \left |\frac{1-A^2\mathrm{e}^{\mathrm{i}\phi_{31}}}{1-A^2\mathrm{e}^{\mathrm{i}\phi_{23}}}\right |^2 + A^2\frac{(1-\mathrm{e}^{\mathrm{i}\phi_{12}})^2}{1-2A^2+A^2\mathrm{e}^{\mathrm{i}\phi_{12}}}\ln \frac{1-A^2\mathrm{e}^{\mathrm{i}\phi_{31}}}{1-A^2\mathrm{e}^{\mathrm{i}\phi_{23}}}-\nonumber \\
&&-\left [\mathrm{e}^{\mathrm{i}\phi_{12}}\ln \frac{1-A^2\mathrm{e}^{\mathrm{i}\phi_{31}}}{1-A^2\mathrm{e}^{\mathrm{i}\phi_{23}}}+ \ln \frac{1-A^2\mathrm{e}^{-\mathrm{i}\phi_{31}}}{1-A^2\mathrm{e}^{-\mathrm{i}\phi_{23}}}\right ]\ .
\label{terms}
\end{eqnarray}
The expansions of all terms except the first one starts with $A^4\equiv A_q^4$ and, consequently,  their sum over $q$ converge for large primes. In the divergent part, as above, we use \eqref{zeta}, and
\begin{equation}
\sum_{q<p^*} \ln \left |\frac{1-A_q^2\mathrm{e}^{\mathrm{i}e_{31}\ln q}}{1-A_q^2\mathrm{e}^{\mathrm{i}e_{23}\ln q}}\right |^2=
\ln  \left |\prod_{q<p^*} \frac{1-A^2\mathrm{e}^{\mathrm{i}e_{31}\ln q}}{1-A^2\mathrm{e}^{\mathrm{i}\phi_{23}\ln q}}\right |^2 
\underset{p^*\to\infty}{\longrightarrow} \ln \left |\frac{e_{23}\zeta(1+\mathrm{i} e_{23})}{e_{31}\zeta(1+\mathrm{i} e_{31})}\right |^2\ .
\end{equation}    
The other terms in \eqref{terms} can be transform as follows
\begin{eqnarray}
&&\frac{\partial}{\partial e_3}\left [\mathrm{e}^{\mathrm{i}\phi_{12}}\ln\frac{1-A^2\mathrm{e}^{\mathrm{i}\phi_{31}}}{
1-A^2\mathrm{e}^{\mathrm{i}\phi_{23}}}+ \ln\frac{1-A^2\mathrm{e}^{-\mathrm{i}\phi_{31}}}{1-A^2\mathrm{e}^{-\mathrm{i}\phi_{23}}}\right ]\nonumber \\
&&=\mathrm{i}\ln q\  A^4\left [ \frac{\mathrm{e}^{-\mathrm{i}\phi_{23}}(\mathrm{e}^{\mathrm{i}\phi_{32}}-\mathrm{e}^{\mathrm{i}\phi_{31}})}
{(1-A^2\mathrm{e}^{\mathrm{i}\phi_{32}})(1-A^2\mathrm{e}^{\mathrm{i}\phi_{31}})} +\frac{\mathrm{e}^{-\mathrm{i}\phi_{31}}(\mathrm{e}^{-\mathrm{i}\phi_{31}}-\mathrm{e}^{-\mathrm{i}\phi_{32}})}{(1-A^2\mathrm{e}^{-\mathrm{i}\phi_{31}})(1-A^2\mathrm{e}^{-\mathrm{i}\phi_{32}})}\right ]\ .
\end{eqnarray}
Similarly, the other terms in Eq.~(\ref{terms}) takes the form 
\begin{eqnarray}
&&\frac{\partial}{\partial e_3}\left [ 
A^2\frac{(1-\mathrm{e}^{\mathrm{i}\phi_{12}})^2}{1-2A^2+A^2\mathrm{e}^{\mathrm{i}\phi_{12}}}\ln\frac{1-A^2\mathrm{e}^{\mathrm{i}\phi_{31}}}{1-A^2\mathrm{e}^{\mathrm{i}\phi_{23}}}\right ]=\nonumber \\
&&=-\mathrm{i}\ln q \ A^4\frac{(1-\mathrm{e}^{\mathrm{i}\phi_{12}})^2}{1-2A^2+A^2\mathrm{e}^{\mathrm{i}\phi_{12}}}\left (\frac{\mathrm{e}^{\mathrm{i}\phi_{31}}}{1-A^2\mathrm{e}^{\mathrm{i}\phi_{31}}} +    \frac{\mathrm{e}^{-\mathrm{i}\phi_{32}}}{1-A^2\mathrm{e}^{-\mathrm{i}\phi_{32}}}  \right )\ .
\end{eqnarray}
Combining all terms together one finds
\begin{eqnarray}
&&R_3^{\mbox{osc}}(e_1,e_2,e_3)=
-\frac{\mathrm{e}^{2\pi \mathrm{i}\overline{d(E)} e_{12}}}{(2\pi \mathrm{i})^3}
|\zeta(1+\mathrm{i}e_{12})|^2
\prod_p\left (1 -\frac{(1-p^{\mathrm{i}e_{12}})^2}{(p-1)^2}\right )  
 \left [\frac{\partial}{\partial e_3}
\ln\left |\frac{\zeta(1+\mathrm{i}e_{32})}{\zeta(1+\mathrm{i}e_{31})}\right |^2 \right . \nonumber\\
&&\left . -\mathrm{i} \sum_q \ln q   \left ( \frac{ q^{\mathrm{i} e_{12}}-1}
{(q^{1+\mathrm{i} e_{23}}-1)(q^{1+\mathrm{i} e_{13}} -1)}\right . \right . +\frac{q^{\mathrm{i} e_{12}}-1}{(q^{1+\mathrm{i} e_{31}}-1)(q^{1+\mathrm{i} e_{32}}-1)}+\label{rosc}\\
&&+ \left . \left .\frac{(1-q^{\mathrm{i} e_{12}})^2}{q-2+q^{\mathrm{i} e_{12}}} 
(\frac{1}{q^{1-\mathrm{i}e_{31}}-1}+\frac{1}{q^{1-\mathrm{i}e_{23}}-1}) \right )\right ]+\mathrm{cyclic\; permutations} 
+\mathrm{c.c.}\ .
\nonumber
\end{eqnarray}
Here "cyclic permutations" means that one has to add 2 other terms corresponding to cyclic permutations of indices $(1,2,3)$, i.e. 
terms with substitutions: $1\to 3,\; 2\to 1,\; 3\to 2$ and  $1\to 2,\; 2\to 3,\; 3\to 1$.

\section{Summary}\label{summary} 

The principal ingredients of the proposed method for calculating correlation functions for the Riemann zeros are the following: 
\begin{itemize}
\item A 'universal' formula for the kernel of a GUE-type ensemble of random matrices with a given mean eigenvalue density, $\bar{\rho}(E)$, 
\begin{equation}
K(x,y)=\frac{\sin \pi \int_{x}^y \bar{\rho}(E)\mathrm{d}E }{\pi (x-y)}\ .
\end{equation}
\item The relation with the Riemann zeta function is established by fixing $\bar{\rho}(E)$ in the above expression  as  the finite part of the density of the zeros
\begin{equation}
\bar{\rho}(E)=\frac{1}{2\pi}\ln \frac{E}{2\pi}  -\frac{1}{2\pi}\sum_{p<p*}\sum_{n=1}^{\infty} 
\frac{\ln p}{ p^{n/2}}\left (\mathrm{e}^{\mathrm{i} n E\ln p}+\mathrm{e}^{-\mathrm{i} n E\ln p}\right ) 
\end{equation}
where the summation is performed over all prime numbers up to a certain cut-off value of $p^*$.
\item Correlation functions are calculated by the averaging the GUE determinantal formula over a large window of $E$
\begin{equation}
 R(e_1,\ldots, e_n)=\left \langle \left \langle \det( K (x,y) )_{x,y=E+e_1,\ldots, E+e_n}\right \rangle \right\rangle_{\Delta E}
 \ .
\end{equation}
\item The average is assumed to be such that phases $E\ln p$ with different primes $p<p^*$ can be considered as independent random phases and the procedure of averaging is carried out by integration over these phases
\begin{equation}
\left \langle \left \langle F\Big (\mathrm{e}^{\mathrm{i}E\ln p_1},\ldots, \mathrm{e}^{\mathrm{i}E\ln p_n} \Big ) \right \rangle \right\rangle_{\Delta E}
=\int_0^{2\pi}\frac{\mathrm{d}\phi_1}{2\pi}\cdots \int_0^{2\pi}\frac{\mathrm{d}\phi_n}{2\pi}
F\Big (\mathrm{e}^{\mathrm{i}\phi_1},\mathrm{e}^{\mathrm{i}\phi_2},\ldots, \mathrm{e}^{\mathrm{i}\phi_n} \Big )\ .
\end{equation}
\item After the averaging,  the result consists of different sums over prime less than $p^*$. Those sums which converge when $p^*\to\infty$ are substituted by the sums over all primes.   Sums divergent at large  $p^*$ can be transformed to one particular product (or its logarithm or its derivative) whose limiting value is
\begin{equation}
\prod_{p<p^*}\dfrac{1-p^{-1} }{1-p^{-1-s}}\underset{p^*\to\infty}{\longrightarrow} s\zeta(1+s)\ . 
\end{equation}
\end{itemize}
When the above rules are accepted, the calculation of correlation functions of Riemann zeros at a large height $E$ on the critical line reduces to purly algebraic manipulations (cf. Sections \ref{two_point} and \ref{three_point}).   Our purpose here was to explain this new method, which, as we have emphasized, has the advantage over other heuristic approaches that it incorporates the determinanetal structure of RMT at the beginning (usually, it requires delicate combinatorial manipulations to establish this \cite{nonlinearity, RudSar, Zeev, CS}).  The method also extends straightforwardly to similar quantum chaotic problems.

For convenience we rewrite the obtained expressions given by  Eqs.~(\ref{rthree}), (\ref{rosc}), and (\ref{rdiag}) together with Eqs.~\eqref{rtwo}, \eqref{r_2diag}, \eqref{r_2osc}  for the two-point and three-point correlation functions.

\begin{center} \textbf{Two-point correlation function} \end{center}

\begin{equation}
R_2(\epsilon)=\overline{d(E)}^{\, 2}+R_2^{\mbox{c}}(\epsilon),\qquad 
R_2^{\mbox{c}}(\epsilon)=R_{2}^{\mbox{diag}}(\epsilon)+R_{2}^{\mbox{osc}}(\epsilon)\ .
\end{equation}
where
\begin{eqnarray}
R_{2}^{\mbox{diag}}(\epsilon)&=&-\frac{1}{4\pi^2}\frac{\partial^2}{\partial \epsilon^2}
\ln|\zeta(1+\mathrm{i}\epsilon)|^2 -
\frac{1}{4\pi^2}\sum_p \ln^2 p \left (\frac{1}{(p^{1+\mathrm{i}\epsilon}-1)^2}
+\frac{1}{(p^{1-\mathrm{i}\epsilon}-1)^2}\right )\ ,
\label{twodiag}\\
R_{2}^{\mbox{osc}}(\epsilon)&=&\frac{\mathrm{e}^{2\pi i \overline{d(E)}\epsilon}}{4\pi^2}|\zeta(1+\mathrm{i}\epsilon)|^2
\prod_p\left (1-\frac{(1-p^{\mathrm{i}\epsilon})^2}{(p-1)^2}\right ) + \mbox{ c.c.}\ .
\label{twoosc}
\end{eqnarray}

\begin{center} \textbf{Three-point correlation function} \end{center}

\begin{eqnarray}
R_3(e_1,e_2,e_3)&=&\overline{d(E)}^{\, 3}+\overline{d(E)}R_2^{\mbox{c}}(e_{12})+
\overline{d(E)} R_2^{\mbox{c}}(e_{23})+  
\overline{d(E)} R_2^{\mbox{c}}(e_{31})+R_3^{\mbox{c}}(e_1,e_2,e_3)\ , \nonumber\\
R_3^{\mbox{c}}(e_1,e_2,e_3)&=& R_{3}^{\mbox{diag}}(e_1,e_2,e_3)+R_{3}^{\mbox{osc}}(e_1,e_2,e_3)\ ,
\end{eqnarray}
where
\begin{eqnarray}
&&R_{3}^{\mbox{diag}}(e_1,e_2,e_3)= -\frac{1}{(2\pi)^3}
\sum_p \ln^3  p\left (
\frac{1}{(p^{1-\mathrm{i} e_{12}}-1)(p^{1-\mathrm{i} e_{13}}-1)}\right . +  
\frac{1}{(p^{1-\mathrm{i} e_{21}}-1)(p^{1-\mathrm{i} e_{23}}-1)}\nonumber\\
&& \left . +\frac{1}{(p^{1-\mathrm{i} e_{32}}-1)(p^{1-\mathrm{i} e_{31}}-1)}\right ) +\mbox{c.c.} 
\end{eqnarray}
and 
\begin{eqnarray}
&&R_3^{\mbox{osc}}(e_1,e_2,e_3)=-\frac{\mathrm{e}^{2\pi \mathrm{i}\overline{d(E)}e_{12}}}{(2\pi \mathrm{i})^3}
|\zeta(1+\mathrm{i}e_{12})|^2
\prod_p \left (1 -\frac{(1-p^{\mathrm{i}e_{12}})^2}{(p-1)^2}\right ) \left [
\frac{\partial}{\partial e_3}\ln
\left |\frac{\zeta(1+\mathrm{i}e_{32})}{\zeta(1+\mathrm{i}e_{31})}\right |^2 \right .\nonumber\\
&&\left .-\mathrm{i} \sum_p \ln p \left ( \frac{ p^{\mathrm{i} e_{12}}-1}
{(p^{1+\mathrm{i} e_{23}}-1)(p^{1+\mathrm{i} e_{13}}-1)}\right .\right .+\frac{p^{\mathrm{i} e_{12}}-1}{(p^{1-\mathrm{i} e_{13}}-1)(p^{1-\mathrm{i} e_{22}}-1)}+
\label{threeosc}\\
&&+ \left . \left .\frac{(1-p^{\mathrm{i} e_{12}})^2}{p-2+p^{\mathrm{i} e_{12}}} 
\Big (\frac{1}{p^{1-\mathrm{i}e_{31}}-1}+\frac{1}{p^{1-\mathrm{i}e_{23}}-1}\Big ) \right )\right ]+
\mathrm{ cyclic \;permutations}+\mathrm{c.c.}\ .
\nonumber
\end{eqnarray}


\end{document}